# Surface Treatment of Cu:NiOx Hole-Transporting Layer Using β-Alanine for Hysteresis-Free and Thermally Stable Inverted Perovskite Solar Cells

**Fedros Galatopoulos, Ioannis T. Papadas, Apostolos Ioakeimidis, Polyvios Eleftheriou and Stelios A. Choulis \***

Molecular Electronics and Photonics Research Unit, Department of Mechanical Engineering and Materials Science and Engineering, Cyprus University of Technology, 3603 Limassol, Cyprus
\* Correspondence: stelios.choulis@cut.ac.cy

**Abstract:** Inverted perovskite solar cells (PSCs) using a Cu:NiOx hole transporting layer (HTL) often exhibit stability issues and in some cases J/V hysteresis. In this work, we developed a β-alanine surface treatment process on Cu:NiOx HTL that provides J/V hysteresis-free, highly efficient, and thermally stable inverted PSCs. The improved device performance due to β-alanine-treated Cu:NiOx HTL is attributed to the formation of an intimate Cu:NiOx/perovskite interface and reduced charge trap density in the bulk perovskite active layer. The β-alanine surface treatment process on Cu:NiOx HTL eliminates major thermal degradation mechanisms, providing 40 times increased lifetime performance under accelerated heat lifetime conditions. By using the proposed surface treatment, we report optimized devices with high power conversion efficiency (PCE) (up to 15.51%) and up to 1000 h lifetime under accelerated heat lifetime conditions (60 °C, $N_2$).

## 1. Introduction

Perovskite solar cells (PSCs) have undoubtedly been the center of attention amongst the most promising photovoltaic (PV) technologies during recent years. This technology has shown great promise over the years with a respectable improvement in power conversion efficiency (PCE) from 3.8 to 25.2% [1,2]. The high PCE of PVSCs can be attributed to several key features that are extremely attractive for PV technologies such as tunable band gap [3,4], low exciton binding energy [5,6], long carrier diffusion length [7], high carrier mobility [8], and high absorption coefficient [9]. Equally important is the compatibility of PVSCs with low temperature solution-processed fabrication techniques that also render them very cost-effective [10].

Despite their high PCE, PVSCs exhibit relatively low stability in environmental conditions, such as heat, moisture, and light, which still renders the commercialization of such devices very challenging. Humidity can rapidly decompose the perovskite crystal due to the hygroscopic nature of its cations such as methylammonium (MA), resulting in the formation of $PbI_2$. Several approaches have been reported in the literature to shield PSCs from moisture such as using hydrophobic thiols [11] and interlayers such as cerium oxide (CeOx) [12]. Replacement of MA with less hydroscopic cations to form mixed cation PSCs (such as $FA_{0.83}Cs_{0.17}Pb(I_{0.8}Br_{0.2})_3$) was also reported to improve the intrinsic stability of PVSCs [13]. Ultraviolent illumination in conjunction with atmospheric oxygen was also reported to negatively affect the intrinsic stability of PVSCs due to the formation of super oxide ions ($O_2^-$). $O_2^-$ can deprotonate the MA cation and cause decomposition of the perovskite crystal [14,15]. Reports in the literature have also reported improved light stability by doping the [6,6]-phenyl-butyric acid methyl ester (PCBM) with graphene quantum dots (GQDs) [16]. Heat is another environmental factor that can negatively affect the stability of PSCs, especially when it is combined with ambient conditions [15]. Using nickel oxide (NiO) and PCBM as the hole transporting layer (HTL) and electron transporting layer (ETL), respectively, has been reported to improve the thermal stability of PVSCs [17]. Apart from the intrinsic stability of the PSC active layer, electrodes and interfaces play an important role as well in stability. For instance, it has been shown that by incorporating a thin Cr layer between the HTL and PSK active layer, diffusion of Au from the metal electrode into the active layer can be reduced [18]. In our most recent work, we also showed that a top electrode using $\gamma$-$Fe_2O_3$ ($PC_{70}BM/\gamma$-$Fe_2O_3$/Al) improved the properties and stability of inverted PSCs under accelerated heat conditions [19].

Amongst the various metal oxides that have been used as HTLs, NiOx has been a popular choice for PVSCs due to its transparency and favorable energy level alignment, which enables improved PCE, particularly due to the improvement of $V_{oc}$ and $J_{sc}$ [20–22]. Nickel oxide has a rock salt structure and exhibits p-type conductivity with a wide bandgap between 3.5 and 4.0 eV while being transparent to visible light. Due to its poor conductivity, doping with several metals such as Li, Mg, Ag, and particularly Cu is very common practice for improvement [20]. Cu is a particularly popular dopant due to having a similar ionic radius with Ni (~0.69 Å) [23]. Solution Combustion-synthesized (SCS) Cu:NiOx has been reported to have a conductivity value of $1.25 \times 10^{-3}$ S/cm [24]. An alternative approach to improve the conductivity of NiOx as well as optimizing its hole collection properties and interface with the perovskite has been reported by Lin et al. In their work they employed a two-step annealing process to minimize issues associated with high temperature annealing. In their report, they showed improved conductivity and PCE as well as improved ambient stability [25] .Due to its chemical stability in ambient conditions and low cost, NiOx has been used as a HTL in several recent PSCs applications. More recently, NiOx has been used as the HTL during an ambient perovskite film fabrication that utilized a prenucleation strategy based on water induced fast intermediate crystallite growth, resulting in highly compact perovskite films [26]. In general, the high robustness, cheap fabrication, suitable energy level alignment and easy processing in ambient conditions render NiOx a very attractive alternative choice for HTL within inverted PSC device structures compared to its organic counterparts such as PEDOT:PSS and PTAA [27].

Although NiOx has been widely used as an effective HTL in terms of both PCE and stability; its behavior has also been reported to be inconsistent. In particular, the stability and performance of NiOx-based devices can vary based on the deposition methods of the NiOx films. For instance, PSCs based on combustion synthesis (CS)of NiOx films have been reported to exhibit better ambient stability compared to their sol-gel counterparts due to better interface with the perovskite as well as better perovskite film quality [28]. A more recent approach for Cu:NiOx synthesis incorporates a sol-gel approach with various organic solvents in order to avoid toxic catalysts [29]. Another factor that can negatively impact the stability of PSCs is the presence of organic ligands on the surface of nanoparticle suspensions of NiOx [30]. Even by solely considering the nature of NiOx films, the situation is not straightforward. The PCE of NiOx-based PSCs was reported to have a large variation between 8% and 20%, even in cases where similar synthesis and deposition methods of NiOx were used [29]. Furthermore, hysteresis effects have also been observed. It has been reported in the literature that such effects can manifest due to the composition of NiOx, which consists of NiO and $Ni_2O_3$, from which NiO is known to have strong dipole moment [31]. This can cause device limitations due to the appearance of surface dipoles in NiOx, which can attract perovskite precursor ions (e.g., $CH_3NH_3^+$, $I^-$, and $Pb^{2+}$) during the crystallization process and induce ionic vacancies and defects in the perovskite crystal. This is particularly apparent if NiOx is used as a HTL for inverted p-i-n PVSCs, because it acts as the underlayer and will affect the crystallization process of the perovskite [29].

Metal oxides interact with atmospheric oxygen, especially when photoexcited by UV light. Different from most organic semiconductors, which will become oxidized, metal oxides can catalyze oxidative decomposition of other organic materials that they come in contact with. Oxygen is readily adsorbed at oxygen vacancies in metal oxides, where it can form reactive superoxide species upon UV light exposure [32]. These superoxide species will ultimately oxidize most materials they come in contact with, including metal halide perovskites. In this study, we propose the application of organic interfacial modifiers such as β-alanine (3-aminopropanoic acid) with the chemical formula of $NH_2CH_2CH_2COOH$), as a surface treatment method for metal oxides to eliminate oxygen vacancies, avoiding interfacial recombination and enhancing device stability [33]. The amino acid carboxyl groups have been previously shown to be the dominant anchoring points on surface treatment applications for metal oxides utilizing mechanisms such as electrostatic attraction, H-bonding, monodentate (ester-like linkage), bidentate bridging, and bidentate chelating [34].

Amino acids have been used before in the field of PSCs either as cross-linkers [35] or anchoring groups [36] due to their unique structure ($^-HOOC-R-NH_3^+$). In terms of cross-linking, their structure allows them to form hydrogen bonds between the cations and halogen ions of the perovskite crystal, thus improving its properties [35]. Furthermore, they can be adsorbed in the surface of metal oxides through their $COOH^-$ group, leaving their $NH_3^+$ group exposed, which in return can contribute to the crystallization process of the perovskite[36]. A common amino acid is β-alanine, which has been previously used in the form of β-alanine-HI salt, acting as an anchoring group on the surface of porous $TiO_2$, improving the PCE of normal-structured PSCs [36]. β-Alanine has also been used as an additive to PEDOT:PSS in order to reconstruct the distribution of $–SO_3^-$ and $–SO_3H$, thus improving charge transport and collection as well as its hydrophobicity [37].

In this work, we developed a β-alanine-based surface treatment for Cu:NiOx HTLs for the inverted (p-i-n) PVSCs device structure. Due to a combination of a more intimate Cu:NiOx/Perovskite interface and passivation of oxygen vacancies, we achieved reduced charge trap density both at the interface as well as in the bulk perovskite layer. Using the proposed β-alanine surface treatment, we significantly reduced the hysteresis of Cu:NiOx-based PVSCs and achieved a champion PCE of 15.51% while also improving stability from 24 to 1000 h under accelerated heat conditions in an inert atmosphere (60 °C, $N_2$).

## 2. Materials and Methods

**Materials:** Toluene (99.7%), nickel (II) nitrate hexahydrate (>98.5%) ($Ni(NO_3)_2·6H_2O$,) copper (II) nitrate trihydrate (99%–104%) ($Cu(NO_3)_2·3H_2O$), 2-methoxyethanol anhydrous (99.8%), acetylacetone (≥99%), β-alanine (99%), dimethyl sulfoxide (≥99.7%),γ-butyrolactone (≥99%), and chlorobenzene anhydrous (99.8%) were purchased from Sigma-Aldrich Chemical. Methylammonium iodide (≥99%) was purchased from GreatCell Solar. Lead (II) iodide (99.999%) and bathocuproin (98%) were purchased from Alfa Aeser. Phenyl-C70-butyric acid methyl ester (99%) was purchased from Solenne BV. Cu pellets were purchased from Kurt J. Lesker. Ultrapure water was produced by a milli-Q Academic system, Millipore (Burlington, MA, USA). All solutions were prepared with analytical grade chemicals and ultrapure milli-Q water with a conductivity of 18.2 μS/cm. ITO patterned glass substrates (sheet resistance 4 Ω/sq.) were purchased from Psiotec Ltd (Berkhamsted, UK).

**Cu:NiOx solution:** Combustion synthesis of Cu:NiOx, 0.95 mmol of $Ni(NO_3)_2·6H_2O$ and 0.05 mmol of $Cu(NO_3)_2·3H_2O$ were dissolved in 2.5 mL 2-methoxyethanol. The solutions were stirred at 50 °C for 1 h; then, 0.1 mmol acetylacetone was added to the solution, and the whole solution was left for further stirring for 1 h at room temperature.

**β-Alanine solution:** Different concentrations of β-alanine (30, 10, 8, and 6 mg/mL) were prepared using deionized $H_2O$, and the pH was adjusted to 4.2 with 1M $HNO_3$. The solution was stirred at room temperature for 30 min.

**Perovskite solution:** The perovskite solution was prepared using a mixture MAI:PbI$_2$ (1:1). The mixture was dissolved in a solvent of γ-butyrolactone:DMSO (7:3). The solution was stirred at 60 °C for 1 h. The perovskite solution was left to cool at room temperature inside the glove box followed by filtering using a 0.22 μm PVDF filter.

**PC$_{70}$BM solution:** The PC$_{70}$BM solution was prepared in a concentration of 20 mg/mL in chlorobenzene. The solution was left over night at 60 °C.

**Device fabrication:** ITO-patterned glass substrates were cleaned using an ultrasonic bath for 10 min in acetone followed by 10 min in isopropanol. The Cu:NiOx films were coated using Doctor Blade with a blade speed of 5 mm/s and a plate temperature of 85 °C. The films were annealed at 300 °C on a hot plate for 1 h in ambient atmosphere. For the surface treatment, the β-alanine solution was drop-casted on the Cu:NiOx films which were left to adhere for 5 min. Following that step, the spin coating process was initiated on the Cu:NiO$_x$ layer at 6000 rpm for 40 sec. The β-alanine-treated films were annealed at 100 °C for 10 min. The perovskite films were coated inside a N$_2$ atmosphere glovebox using a three-step spin-coating process: first step—500 rpm for 5 s; second step—1000 rpm for 45 s; and third step—5000 rpm for 45 s. During the third step, after the first 20 s of the duration of the step, 0.5 mL toluene was dropped onto the spinning substrate as the anti-solvent to achieve the rapid crystallization of the films. The resulting perovskite films were annealed at 100 °C for 10 min. The PC$_{70}$BM film was coated inside the glovebox using spin coating at 1300 rpm for 30 s. The BCP and Cu layers were deposited using thermal evaporation.

**Device Characterization:** The thicknesses and the surface profile of the device layers were measured with a Veeco Dektak 150 profilometer (Hong Kong, China). The current density-voltage (J/V) characteristics were characterized by a Botest LIV Functionality Test System (Kreuzwertheim, Germany). Both forward (short circuit-open circuit) and reverse (open circuit-short circuit) scans were measured with 10 mV voltage steps and 40 ms of delay time. For illumination, a calibrated Newport Solar simulator (Irvine, CA, USA) equipped with a Xe lamp was used, providing an AM1.5G spectrum at 100 mW/cm$^2$ as measured by a certified oriel 91,150 V calibration cell. A custom-made shadow mask was attached to each device prior to measurements to accurately define the corresponding device area (9 mm$^2$). External quantum efficiency (EQE) measurements were performed by a Newport System (Irvine, CA, USA), Model 70356_70316NS. Impedance spectroscopy was performed using a Metrohm Autolab PGSTAT 302N equipped with FRA32M module (Herisau, Switzerland). To extract the Nyquist plots, the devices were illuminated using a white light emitting diode (LED). A small AC perturbation voltage of 10 mV was applied, and the current output was measured at a frequency range of 1 MHz to 1 Hz. The steady-state DC bias was kept at 0 V. Mott–Schottky measurements were using a Metrohm Autolab PGSTAT 302N potentiostat. The Mott–Schottky measurements were performed using a Metrohm Autolab PGSTAT 302N potentiostat by connecting the anode and the cathode of the PSC devices directly to the system. The frequency values used were 5 and 100 kHz with a bias range of −0.5–1.3 V under dark. Photoluminescence (PL) spectra were obtained at room temperature on a Jobin-Yvon Horiba FluoroMax-P (SPEX) spectrofluorometer (Singapore) equipped with a 150 W Xenon lamp and operated from 300 to 900 nm. Transmittance and absorption measurements were performed with a Shimadzu UV-2700 UV–vis spectrophotometer (Kyoto, Japan). Atomic force microscopy (AFM) data were obtained using Nanosurf easy scan 2 controllers in tapping mode. The contact angle measurements were performed using a Kruss drop shape analyzer equipped with DSA 100 camera. The conductivity measurements were performed using a Jandel RM3000 four-point probe system. The aging of devices was conducted in a hotplate at 60 °C inside a N$_2$ glovebox atmosphere. The devices were taken outside the glovebox in order to measure their PV parameters and undergo device characterization when it was needed at 24 h intervals. The devices were placed back in the glovebox to continue the aging process every 24 h.

## 3. Results and Discussion

### 3.1. Film Characterization

In order to study the effect of β-alanine-treated Cu:NiOx on the performance of p-i-n PSCs, an initial study was performed on both the film properties of Cu:NiOx and Cu:NiOx/β-alanine as well as the perovskite film crystallized on top of Cu:NiOx and Cu:NiOx accordingly.

Figure 1 shows the AFM data as well as the calculated roughness (RMS) and peak to valley (P-V) of the films. Initially, the RMS of both Cu:NiOx and Cu:NiOx/β-alanine on quartz were calculated and showed similar values at ~3 nm. It is also worth noting that when the perovskite film was coated on top of ITO/Cu:NiOx and ITO/Cu:NiOx/β-alanine, it had similar RMS of ~10 nm. This indicates that β-alanine surface treatment does not introduce any extra roughness to the Cu:NiOx film. Although the RMS of the perovskite film was very similar, there was a distinguishable difference in P-V. In detail, P-V decreased from 90 to 76 nm when the perovskite film crystallized on top of ITO/Cu:NiOx/β-alanine compared to ITO/Cu:NiOx. P-V is defined as the distance between the highest peak and the lowest valley of a topographical image and is desirable to be as low as possible in order to ensure compact, smooth, and pinhole free films [38]. The phase contrast data of ITO/Cu:NiOx/Pvsk and ITO/Cu:NiOx/β-alanine/Pvsk are shown in Figure S1. From Figures 1c,d and Figure S1, we report similar morphology of the crystallized perovskite between pristine and alanine treated Cu:NiOx. Although wetting properties in some cases influence the performance of interfaces, a negligible difference in the contact angle measurements of pristine Cu:NiOx and alanine-treated Cu:NiOx was observed (please see Figure S2), indicating that wetting properties by the proposed surface treatment remain unaffected. Furthermore, having a bottom electrode with high transparency in order to ensure efficient light transmission towards the active layer for the generation of carriers is required. Considering these requirements, we also performed transmittance and absorption measurements in ITO/Cu:NiOx, ITO/Cu:NiOx/β-alanine, ITO/Cu:NiOx/Pvsk, and ITO/Cu:NiOx/β-alanine/Pvsk films. The plots shown in Figure S3 indicate that there are no significant differences between either the transmittance of ITO/Cu:NiOx and ITO/Cu:NiOx/β-alanine or the absorption of the perovskite active layer, crystallized on top of the aforementioned two films.

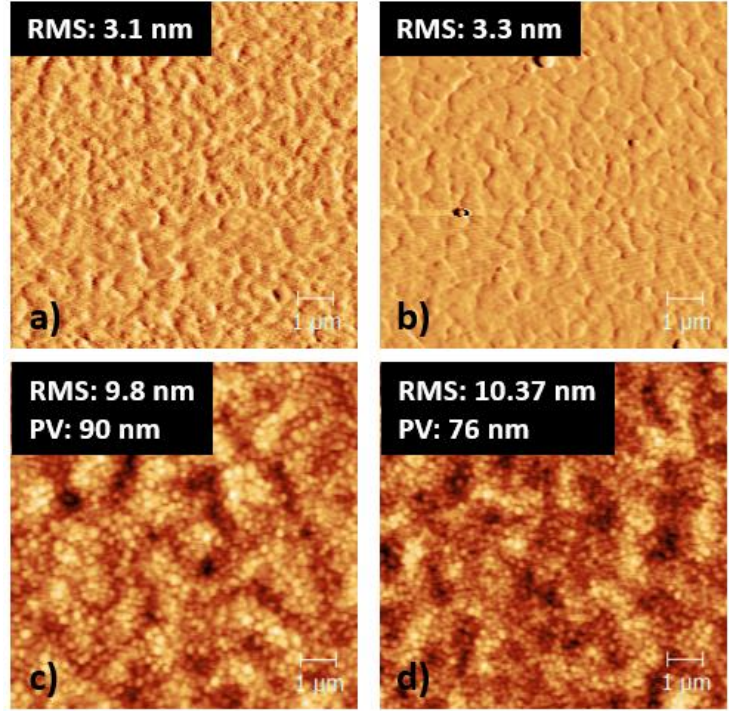

*Figure 1.* AFM images of (**a**) phase contrast of quartz/Cu:NiOx, (**b**) phase contrast of quartz/Cu:NiOx/β-alanine, (**c**) topography of ITO/Cu:NiOx/Perovskite, and (**d**) topography of ITO/Cu:NiOx/β-alanine/Perovskite.

Following the AFM measurements, contact angle measurements, absorption measurements, and photoluminescence (PL) spectroscopy was performed, using an excitation wavelength of 500 nm, in order to determine the charge carrier recombination characteristics of both Cu:NiOx and β-alanine-surface treated Cu:NiOx as well as the perovskite active layer crystallized in the two different films. The absorption spectra shown in Figure 2a show very similar characteristics and we can, therefore, deduce that the band gap ($E_g$) of Cu:NiOx does not change significantly upon β-alanine treatment. The chemical structure of β-alanine that was used for the proposed surface treatment is provided as an inset within Figure 2a [33]. Equally interesting observations are indicated in Figure 2b. ITO/Pvsk exhibited a large peak at 760 nm as expected. Since the HTL was absent, very poor quenching was observed due to the poor carrier extraction and more frequent recombination events. Upon using Cu:NiOx, we immediately observed an improvement to PL(photoluminescence) quenching of more than 90%, indicating a great reduction in the band-to-band charge recombination and a better hole carrier selectivity. Upon β-alanine-treatment of the Cu:NiOx, the PL quenching improved even further (95% quenching) pointing to even more efficient charge collection and an efficient suppression of the electron-hole recombination. The reported experimental results provide indication that charge traps lying in the Cu:NiOx/Pvsk interface are passivated upon treatment with β-alanine. Similar PL quenching behavior was also reported by Shih et al. [39]. In their work, which studied the effect of surface treatment using several amino acids including β-alanine on $TiO_2$, they attributed this observation to the preferential, perpendicular, orientation of the (110) plane of the MAPI crystallites towards the amino acid treated surface. This was observed using grazing-incidence wide-angle X-ray scattering of thin $CH_3NH_3PbI_3$ perovskite films [39]. This resulted in a more intimate interface of the metal oxide with MAPI, reducing the charge trap density of the metal oxide/MAPI interface. We hypothesized that we would observe a similar behavior for the alanine treatment of Cu:NiOx. Furthermore Shih et.al also performed UPS measurements and showed that the $E_{wf}$ of $TiO_2$ upon alanine treatment remains essentially the same [39]. Based on their results, we did not expect a considerable shift to the $E_{wf}$ of Cu:NiOx either upon alanine treatment. Finally, the conductivity values of pristine Cu:NiOx and alanine treated Cu:NiOx films were measured using a four-point probe system, and no considerable change was observed(Cu:NiOx: 4.7 × 10$^{-3}$ S/cm, Cu:NiOx/alanine 3.5 × 10$^{-3}$ S/cm)—in good agreement with the values reported in the literature [24]. The similar conductivity values of Cu:NiOx and alanine treated Cu:NiOx indicated that non-intentional doping effects are unlikely due to the proposed surface treatment.

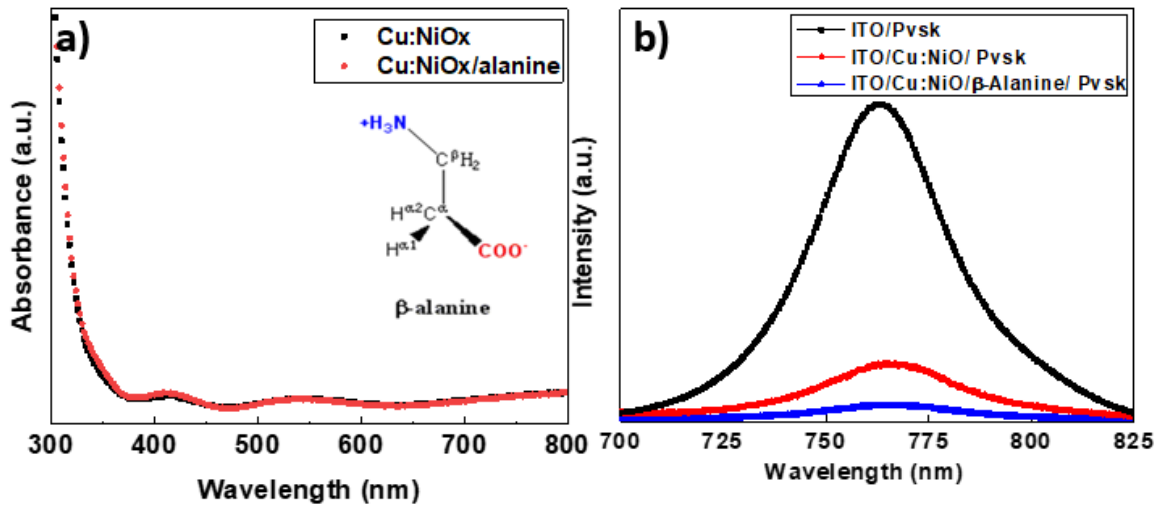

*Figure 2.* (**a**) *Absorption spectra of ITO/Cu:NiOx, ITO/Cu:NiOx/β-alanine and β-alanine structure that used for the surface treatment method (the inset of Figure 2a provides the chemical structure of β-alanine used for the proposed surface treatment).* (**b**) *PL spectra of ITO/Pvsk, ITO/Cu:NiOx/Pvsk, ITO/Cu:NiOx/β-alanine/Pvsk.*

*3.2. Device Characterization*

Following the film characterization, we fabricated full PSC devices based on two different HTLs—pristine Cu:NiOx and β-alanine-treated Cu:NiOx. Two different thicknesses of Cu:NiOx were tested (20 and 40 nm) as well as three different concentrations of β-alanine solutions (30, 10, and 6 mg/mL). The perovskite active layer used was based on a CH$_3$NH$_3$PbI$_3$ (MAPI) formulation, detailed in the materials and methods section. The main device structure used was ITO/Cu:NiOx (40nm)/CH$_3$NH$_3$PbI$_3$ (300 nm)/PC$_{70}$BM (70 nm)/BCP (7 nm)/Cu (80 nm). The thickness of each layer was measured using a profilometer. The PV device parameters are shown in Table 1. Standard deviation (SD) of the PCE is shown in Figure S4.

*Table 1. Perovskite solar cells (PSCs) photovoltaic (PV) parameters.*

| Device Architecture | $V_{oc}$ (V) | $J_{sc}$ (mA/cm²) | FF (%) | PCE (%) |
|---|---|---|---|---|
| ITO/Cu:NiOx (40 nm)/CH$_3$NH$_3$PbI$_3$/PC$_{70}$ BM/BCP/Cu | 0.94 | 16.05 | 81.6 | 12.31 |
| ITO/Cu:NiOx (40 nm)/β-alanine (30 mg/mL)/CH$_3$NH$_3$PbI$_3$/PC$_{70}$BM/BCP/Cu | 1.05 | 14.54 | 48.4 | 7.40 |
| ITO/Cu:NiOx (40 nm)/β-alanine (10 mg/mL)/CH$_3$NH$_3$PbI$_3$/PC$_{70}$BM/BCP/Cu | 1.07 | 17.68 | 70.2 | 13.31 |
| ITO/Cu:NiOx (40 nm)/β-alanine (6 mg/mL)/CH$_3$NH$_3$PbI$_3$/PC$_{70}$BM/BCP/Cu | 1.05 | 18.61 | 74 | 14.46 |
| ITO/Cu:NiOx (20 nm)/CH$_3$NH$_3$PbI$_3$/PC$_{70}$BM/BCP/Cu | 1.03 | 17.19 | 79.5 | 14.13 |
| ITO/Cu:NiOx (20 nm)/β-alanine (10 mg/mL)CH$_3$NH$_3$PbI$_3$/PC$_{70}$BM/BCP/Cu | 1.07 | 20.13 | 71.7 | 15.51 |

In Table 1, it is shown that upon surface treatment with β-alanine, the $V_{oc}$ (open circuit voltage) of Cu:NiOx-based devices increased from 0.94 to 1.05 V; however, the FF (fill factor) dropped dramatically from 81.6% to 48.4%. We observed that by decreasing the concentration of the β-alanine solution, the FF improved. This is to be expected since amino acids in general are not conductive [40], and therefore, a very thin layer should be coated on top of the Cu:NiOx to maintain functional devices. It is worth mentioning that apart from $V_{oc}$, the $J_{sc}$ (short circuit current) of devices also improved upon the surface treatment of Cu:NiOx with β-alanine. As we mentioned in the introduction, the stability of devices is equally important. The different device structures mentioned in Table 1 were subjected to accelerated heat conditions in an inert atmosphere (60 °C, $N_2$) in order to test the effect of β-alanine treatment in the stability of the devices.

By examining the lifetime plots in Figure 3, the difference in stability between Cu:NiOx and β-alanine-treated Cu:NiOx is very apparent. The PCE (power conversion efficiency) of Cu:NiOx (40 nm)-based devices dropped at ~40% even in the first 24 h of aging, primarily due to a drop in $J_{sc}$. On the contrary, β-alanine-treated (10 mg/mL) Cu:NiOx devices retain their PCE and $T_{80}$ (determined as the time needed for a device to reach 80% of its initial PCE) was reached at 1000 h of aging. It is interesting to note that when the concentration of β-alanine decreased to 6 mg/mL, the stability also decreased, with the PCE dropping at 60% of its initial value at 24 h of aging, similar to pristine Cu:NiOx. However, even if the PCE dropped to 60% of its initial value, this was retained even at 1000 h of aging, although it was still lower than the standard $T_{80}$. When the thickness of Cu:NiOx decreased from 40 to 20 nm, there was no significant difference observed in the device lifetime performances. The J/V characteristics of devices were studied to understand the mechanisms behind the degradation of Cu:NiOx devices and β-alanine-treated Cu:NiOx devices. Impedance spectroscopy was also used to understand these mechanisms in more detail. The characterization of devices was performed in Cu:NiOx- (40 nm) and Cu:NiOx (40 nm)/β-alanine (10 mg/mL)-based devices. Both device structures were characterized at a time frame where the degradation phenomena were detectable but did not render the device unmeasurable (24 h for untreated Cu:NiOx-based devices and 1000 h for β-alanine treated Cu:NiOx-based devices).

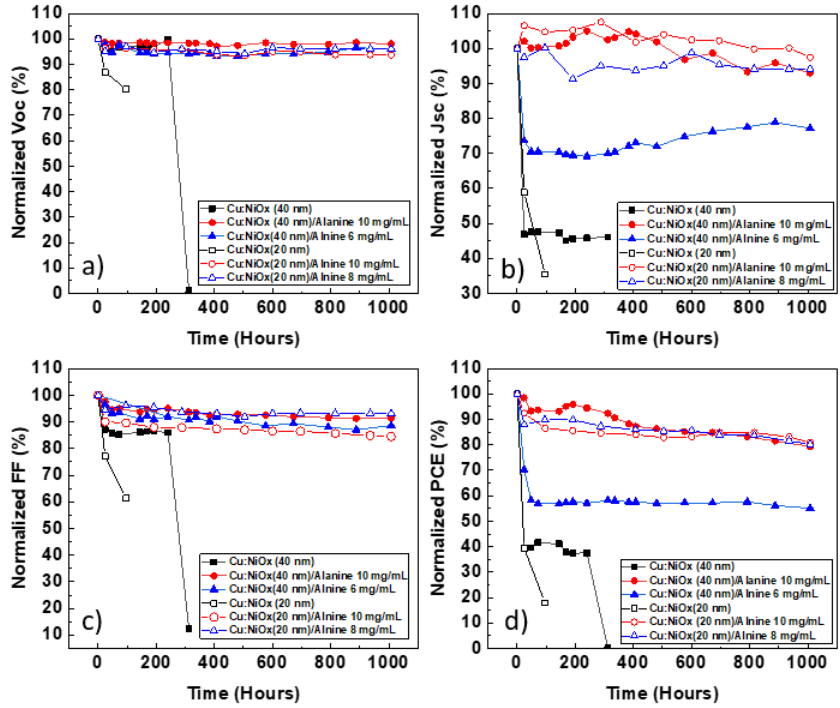

*Figure 3.* Lifetime plots of normalized: (*a*) $V_{oc}$, (*b*) $J_{sc}$, (*c*) FF, and (*d*) PCE under accelerated heat conditions.

Cu:NiOx-based devices exhibited severe hysteresis, as shown in Figure 4a, which was eliminated by leaving the devices under light soaking for 20 min. The PV parameters of Cu:NiOx-based devices, in the forward (FW) and reverse (RV) scan direction, before and after light soaking are shown in Table 2.

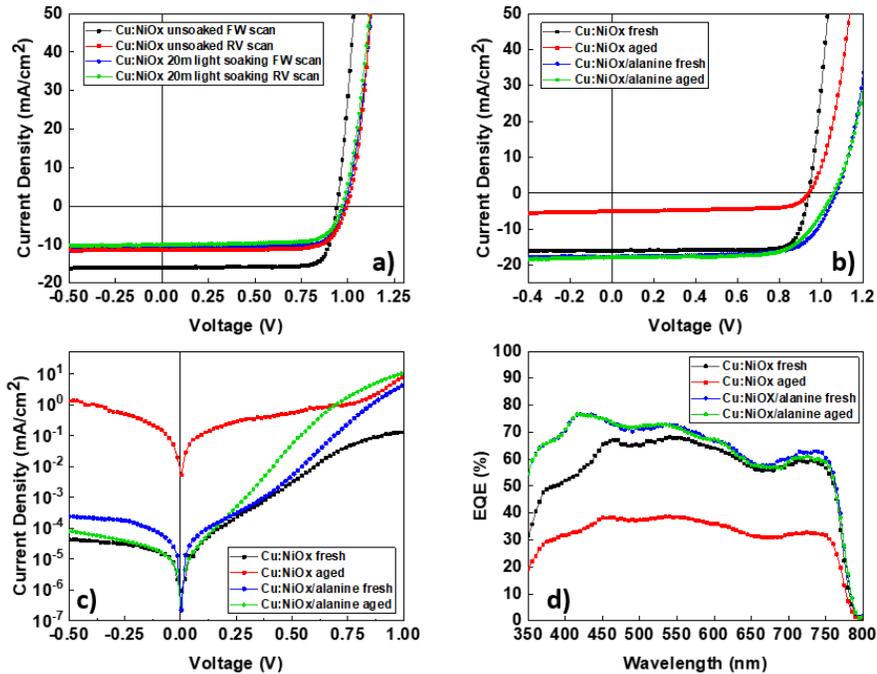

*Figure 4. (a) Illuminated J/V characteristics of Cu:NiOx devices before and after light soaking, (b) illuminated J/V characteristics of fresh and aged Cu:NiOx and β-alanine-treated Cu:NiOx devices, (c) dark J/V characteristics of fresh and aged Cu:NiOx and β-alanine-treated Cu:NiOx devices, (d) EQE spectra of fresh and aged Cu:NiOx and β-alanine-treated Cu:NiOx devices.*

*Table 2. PV parameters of Cu:NiOx-based devices before and after light soaking.*

| Device Architecture | $V_{oc}$ (V) | $J_{sc}$ (mA/cm$^2$) | FF (%) | PCE (%) |
| --- | --- | --- | --- | --- |
| Cu:NiOx-based devices without light soaking | FW = 0.94  RV = 0.99 | FW = 16.05  RV = 11.42 | FW = 81.6  RV = 76.41 | FW = 12.31  RV = 8.61 |
| Cu:NiOx-based devices after 20 m light soaking | FW = 0.98  RV = 0.97 | FW = 10.79  RV = 10 | FW = 77.4  RV = 74.2 | FW = 8.19  RV = 7.20 |

The hysteresis of p-i-n PSC devices based on Ni:Ox was also reported in the literature as we have previously stated, and it was shown to have a recoverable effect on the PCE of the device upon light soaking [29]. In our case, even though the hysteresis reduced significantly after light soaking for 20 min, the PCE was not recovered; on the contrary, it decreased, primarily due to a decrease in $J_{sc}$. Interestingly, the $V_{oc}$ increased after light soaking. J/V hysteresis is a very common phenomenon in PSCs that has been previously ascribed to trapping and de-trapping of charged carriers [41]. Specifically, $J_{sc}$ hysteresis has been previously ascribed to trapping of charged carriers due to bulk defects of the perovskite layer [42]. Upon light soaking, passivation of these traps can occur by the charged carriers themselves; however, less photo generated charges will contribute to the $J_{sc}$ of the device due to a change in local electrical polarization effects on the surface of the crystal, resulting in poorer charge dissociation and reduction in $J_{sc}$. A similar passivation mechanism can be used to explain the increase in $V_{oc}$. $V_{oc}$ is directly related to the built-in electric field ($E_{bi}$) of the device which is affected by the accumulation of charges at the electrodes. The increase in $V_{oc}$ after light soaking can be explained by the reduction in charge accumulation at the interfaces of the devices. Trap passivation in the Cu:NiOx/Pvsk interface due to charged carriers generated after light soaking can cause a reduction in charge accumulation in this interface, thus increasing the $V_{oc}$ [41]. Trap filling phenomena have been previously reported to decrease the FF of PSCs upon light soaking [29]. This is also true in our case where we observed a drop in FF upon light soaking of Cu:NiOx-based devices, as shown in Table 2. Upon surface treatment of Cu:NiOx with β-alanine, we observed an increase in $V_{oc}$ and $J_{sc}$ as well as significant reduction in hysteresis, as shown in Figure 4b. Following the investigation of the observed J/V characteristics of devices with and without light soaking, as well as the $E_{wf}$ data reported from Shih et al. [39], the increase in $V_{oc}$ can be attributed to a reduction in charge traps at the Cu:NiOx perovskite interface and the increase in $J_{sc}$ to reduce bulk defects in the perovskite crystal. The reduction in charge traps both at the interfaces as well as the bulk perovskite layer is also ascribed to the improved stability that β-alanine-treated Cu:NiOx inverted PSCs showed. Significant changes were also observed in the dark J/V characteristics shown in Figure 4c. In detail, β-alanine-treated Cu:NiOx-based devices showed a reduction in the series resistance ($R_s$) from 75.14 to 18.25 Ω and an increase in shunt resistance ($R_{sh}$) from 15.47 to 35.65 kΩ, resulting in the improved $V_{oc}$ and $J_{sc}$ that these devices exhibited. The EQE spectra were also studied and presented in Figure 4d. Interestingly, the shape of the spectra was similar between untreated and β-alanine-treated Cu:NiOx-based devices. The spectra of untreated Cu:NiOx were lower than β-alanine-treated Cu:NiOx-based devices. The EQE spectra of untreated Cu:NiOx-based devices showed a significant decrease upon aging, following the $J_{sc}$ drop, whereas β-alanine-treated Cu:NiOx-based devices showed no significant decrease in their EQE. The photocurrent from the EQE spectra was calculated using the overlap integral of the EQE spectrum with the AM 1.5G global spectrum and was found to have <5% deviation from the calculated $J_{sc}$ using the J/V characteristics for all the device structures under test (Cu:NiOx fresh: EQE current = 15.72 mA/cm$^2$; JV current = 16.05 mA/cm$^2$; Cu:NiOx aged: EQE

current = 8.31 mA/cm$^2$; JV current = 7.91 mA/cm$^2$; Cu:NiOx/alanine fresh: EQEcurrent = 17.76 mA/cm$^2$; JV current = 17.68 mA/cm$^2$; Cu:NiOx/alanine aged: EQE current = 17.47 mA/cm$^2$; JV current = 17.25 mA/cm$^2$).

To obtain a visual representation of the degradation of untreated and β-alanine-treated Cu:NiOx-based devices, we performed photocurrent (PCT) mapping, as shown in Figure 5. PCT allows us to obtain a visual representation of the current distribution in the device. Untreated Cu:NiOx-based devices showed non-uniform current distribution even if they were fresh. There were also areas in the device that showed very little presence of current, which can be tied to the defects. Upon aging, the areas that showed defects were enlarged and show even further current loss. β-alanine-treated devices showed a much more uniform current distribution which was also retained at a large degree even after aging. It is interesting to note that there was a small current loss at the edges of β-alanine-treated devices after aging, as shown in Figure 5d, which can be tied to the ingress of moisture from the edges of the encapsulation cover slip after the prolonged aging test.

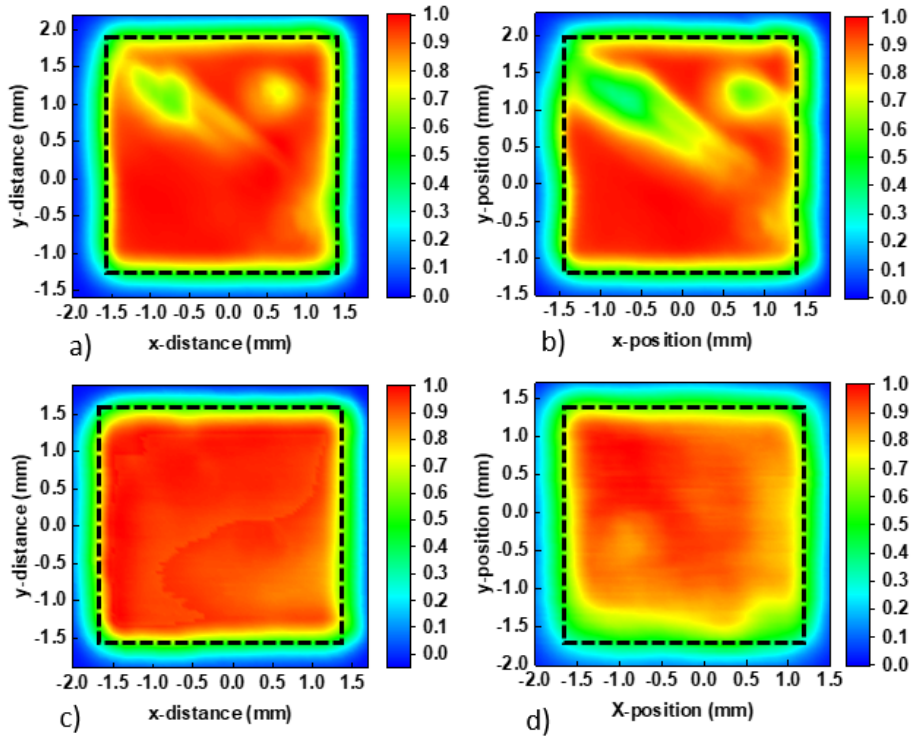

***Figure 5.*** *Photocurrent mapping of (**a**) fresh Cu:NiOx, (**b**) aged Cu:NiOx, (**c**) fresh β-alanine-treated Cu:NiOx ,and (**d**) aged β-alanine-treated Cu:NiOx-based device.*

In order to further investigate the origin of the degradation differences between the two device structures, we performed impedance spectroscopy (IS) and Mott–Schottky (MS) analysis. MS analysis is a very useful characterization technique that can be used to distinguish mechanisms that take place in the active layer with those in the interfaces. In reverse bias, a capacitive plateau appeared that corresponded to the geometric capacitance (Cg), related to the dielectric characteristics and polarizability of the active layer. The point of intersection of the linear slope with the *x*-axis corresponded to the estimated built in voltage (EV$_{bi}$). It is interesting to note that the EV$_{bi}$ deviated from the real V$_{bi}$ value due to the depletion approximation used in the MS analysis. The depletion approximation states that the capacitance in the space charge region manifests purely due to doping and not free charges, which generally does not hold true close to the metal contacts where free charges tend to accumulate. Although there was a deviation between EV$_{bi}$ and V$_{bi}$, it can still be used for comparative studies tied with the V$_{oc}$ [43,44]. By varying the frequency of the AC signal between high (HF) and low frequencies (LF) in MS analysis, we can also detect trapping phenomena. In general, at LF, since we approach DC conditions, trapped charges can de-trap and, therefore, by observing the change in MS plots between HF and LF, we can obtain a rough understanding of the trapping phenomena inside the device [45,46]. In Figure 6a, we can observe an increase in the capacitive plateau from $3.2 \times 10^{-9}$ F/cm$^2$ to $1.4 \times 10^{-8}$ F/cm$^2$ when we changed the frequency from 100 to 5 kHz for untreated Cu:NiOx-based devices. The increase in capacitance at negative biases when we decreased the frequency indicates the presence of trapped charges in the space charge region, which is related to bulk defects of the perovskite layer. Capacitance increases further from $1.4 \times 10^{-8}$ F/cm$^2$ to $7 \times 10^{-7}$ F/cm$^2$ when we aged the untreated Cu:NiOx-based devices and we kept the frequency constant at 5 kHz. When choosing an appropriate frequency for MS analysis, it should usually coincide in the high-frequency plateau of the relevant capacitance frequency (CF) plot [44]. CF measurements were performed using this particular perovskite formulation and a similar p-i-n device structure in our previous work, and therefore, 5 kHz is an appropriate choice for our particular measurements [22]. In comparison, β-alanine-treated Cu:NiOx devices retain their capacitance at ~$2 \times 10^{-9}$ F/cm$^2$ between HF and LF as well as after aging. These results agree with the $J_{sc}$ and hysteresis behavior of both untreated and β-alanine-treated Cu:NiOx-based devices. By observing the Nyquist plots, we can observe two different frequency features at HF and LF, as denoted by the two different semi-circle arcs similar to what is reported in the literature [47]. The HF feature was previously attributed to the charge transport resistance ($R_{tr}$). There is also a small series resistance ($R_s$) shown by the indent of the curve with the *y*-axis that is attributed to the resistance from the experimental setup as well as the ITO [48]. The LF feature was previously attributed to ionic diffusion and the charge recombination resistance ($R_{rec}$) due to charge accumulation at the interfaces[49,50]. Under illumination, the charge accumulation at the interfaces has been reported to be the dominant mechanism, similar to our case [48,51]. Using the equivalent circuit model from our previous work [22], we fitted the Nyquist plot data in order to extract values for $R_{tr}$ and $R_{rec}$ (Figure S5). Upon alanine treatment, we report a drop for $R_{tr}$ from 1.6 to 1.3 kΩ and an increase to $R_{rec}$ from 19.5 to 33 kΩ. These values are in good agreement with the drop of $R_s$ and increase in $R_{sh}$, respectively. Since $R_{rec}$ was tied to the charge accumulation at the interfaces, we can conclude that β-alanine-treated Cu:NiOx devices exhibited reduced charge trap density at the Cu:NiOx/Perovskite interface. This is also in agreement with the increased $EV_{bi}$ in the MS analysis and improved $V_{oc}$ that β-alanine-treated Cu:NiOx devices exhibited.

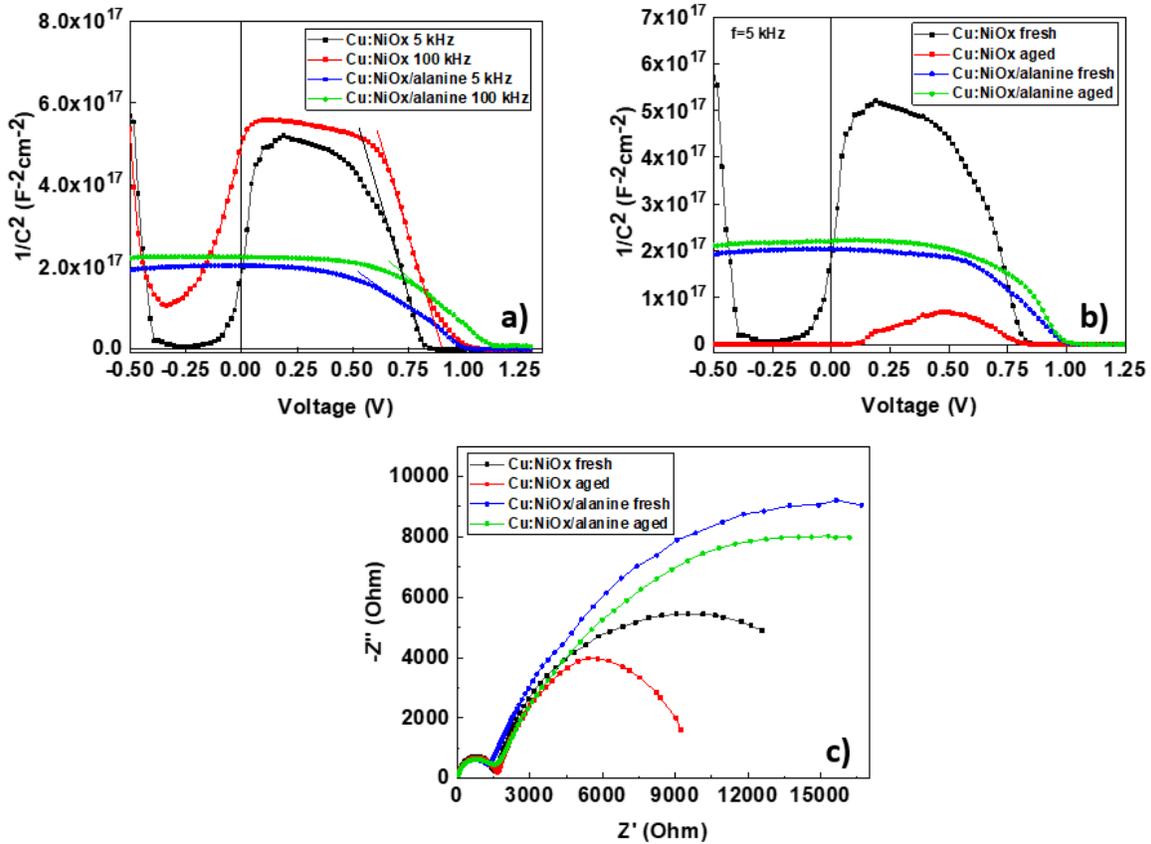

*Figure 6.* Mott–Schottky plots of (**a**) untreated and β-alanine-treated Cu:NiOx-based devices in 5 and 100 kHz, (**b**) fresh and aged untreated an β-alanine-treated Cu:NiOx-based devices at 5 kHz, and (**c**) Nyquist plots fresh an aged untreated and β-alanine-treated Cu:NiOx-based devices.

The presence of traps in the Cu:NiOx/Pvsk interface as well as the degradation mechanisms of the perovskite bulk layer can be explained by examining the nature of metal oxides. In particular, metal oxide functionality, including the Cu:NiOx under investigation within this work, is known to influenced by oxygen vacancies that are caused by the impurities and the changes of stoichiometry [52]. The creation of oxygen defects within the metal-oxide is caused by loss of oxygen atoms which can serve as active sites on which oxidation or reduction take place during photocatalytic activity, or a trapping site which can inhibit and reduce the functionality of inverted PVSCs [52]. Oxygen vacancies can act as trap states themselves by inhibiting the carrier movement inside the charge-selective layer—in our case Cu:NiOx—as well as the Cu:NiOx/Pvsk interface and, therefore, promoting charge recombination in the interfaces of the device. Furthermore, metal oxides are known to form reactive superoxide ($O_2^-$) species by adsorbing atmospheric oxygen through their vacancies and reacting with UV from ambient light. The $O_2^-$ can further negatively impact PVSCs by deprotonating the $MA^+$ cation through oxidation and, therefore, negatively impact both PCE and cause instability by introducing charge traps in the active layer [53]. Deprotonation of the $MA^+$ cation was also recently reported by McGehee et al. to be caused the $Ni^{\geq 3+}$ metal cation sites in NiOx that can act both as Brønsted proton acceptors and Lewis electron acceptors and can react with the amine functional groups. This results in cation-deficient perovskite crystals that cause poor hole extraction in the NiOx/Pvsk interface and promote charge recombination. In their work they used excess cation salts during the Pvsk active layer deposition to neutralize this effect. We believe this happens to an extent in our case as well since alanine can act as a source of amine functional groups, similar to the $MA^+$ cation and, therefore, neutralize the effect described by McGehee et al. [54]. Through the proposed metal-oxide β-alanine treatment, we have shown that the charge trap density can be reduced both at the Cu:NiOx/Pvsk interface as well as the perovskite active layer [52,53]. The experimental results indicate the importance of the reduction in oxygen vacancies and, therefore, passivation of the charge traps for the development of high performance inverted PVSCs that are based on metal-oxide HTLs. By optimizing the thickness of Cu:NiOx, we managed to achieve high PCE while still retaining the improved lifetime. By decreasing the thickness of Cu:NiOx to 20 nm and by using the proposed β-alanine surface treatment, we developed hysteresis-free devices with a PCE = 15.51% (Figure S6) and improved stability with T80 at 1000 h under accelerated heat conditions (60 °C, $N_2$).

## 4. Conclusions

We identified that the main reason for the poor device performance and J/V hysteresis of inverted (p-i-n) PSCs incorporating Cu:NiOx HTLs to be the presence of charge traps at the bulk perovskite active layer as well as the Cu:NiOx/Pvsk interface. We introduced a β-alanine-based metal-oxide surface treatment method that can improve the performance and stability of inverted PVSCs incorporating Cu:NiOx HTLs. Using β-alanine-surface treatment, we reduced interfacial charge trap density due to the formation of a more intimate Cu:NiOx/Pvsk interface, which results in improved $V_{oc}$. The improved $J_{sc}$ is attributed to the reduced oxygen vacancies on β-alanine-treated Cu:NiOx HTL which leads to the reduction in the bulk perovskite active layer charge trap density. Those effects contribute to the improved PCE and stability of inverted PSCs that are using β-alanine-surface treated Cu:NiOx HTLs. Using the β-alanine metal-oxide surface treatment, we reported hysteresis-free inverted PSCs with PCE = 15.51%. More importantly, the thermal stability is greatly improved from 24 h for the untreated Cu:NiOx HTL to 1000 h for the β-alanine-surface treated Cu:NiOx HTL-based inverted PSCs at accelerated heat lifetime conditions in an inert atmosphere (60 °C, $N_2$). We believe that the proposed surface treatment process using β-alanine can be applied in other functional metal oxides, thus paving the way to hysteresis-free, high performance, and long-lived PSCs.

**Supplementary Materials:** The following are available online at www.mdpi.com/xxx/s1, Figure S1: AFM images of: (a) ITO/Cu:NiOx/Pvsk, (b) ITO/Cu:NiOx/β-alanine/Pvsk; Figure S2: contact angle film measurements of (a) Cu:NiOx, (b) Cu:NiOx/alanine; Figure S3: (a) transmittance spectra ITO/Cu:NiOx, ITO/Cu:NiOx/alanine, and (b) absorbance spectra ITO/Cu:NiOx/Pvsk, ITO/Cu:NiOx/β-alanine/Pvsk; Figure S4: average PCE and SD for the optimized devices based on Cu:NiOx (20 nm); Figure S5: (a) Nyquist data and fitting for (a) $R_{tr}$ and (b) $R_{rec}$; Figure S6: J/V characteristic of optimized ITO/Cu:NiOx/β-alanine/Pvsk/PC$_{70}$BM/BCP/Cu.

**Acknowledgments:** This project received funding from the European Research Council (ERC) under the European Union's Horizon 2020 research and innovation program (grant agreement No 647311).

# Supporting Information

# Surface Treatment of Cu:NiOx Hole-Transporting Layer Using β-Alanine for Hysteresis-Free and Thermally Stable Inverted Perovskite Solar Cells

By Fedros Galatopoulos [1], Ioannis T. Papadas [1], Apostolos Ioakeimidis [1], Polyvios Eleftheriou, and Stelios A. Choulis [1*]

[1] Molecular Electronics and Photonics Research Unit, Department of Mechanical Engineering and Materials Science and Engineering, Cyprus University of Technology, Limassol, 3603 (Cyprus). * Corresponding Author: stelios.choulis@cut.ac.cy

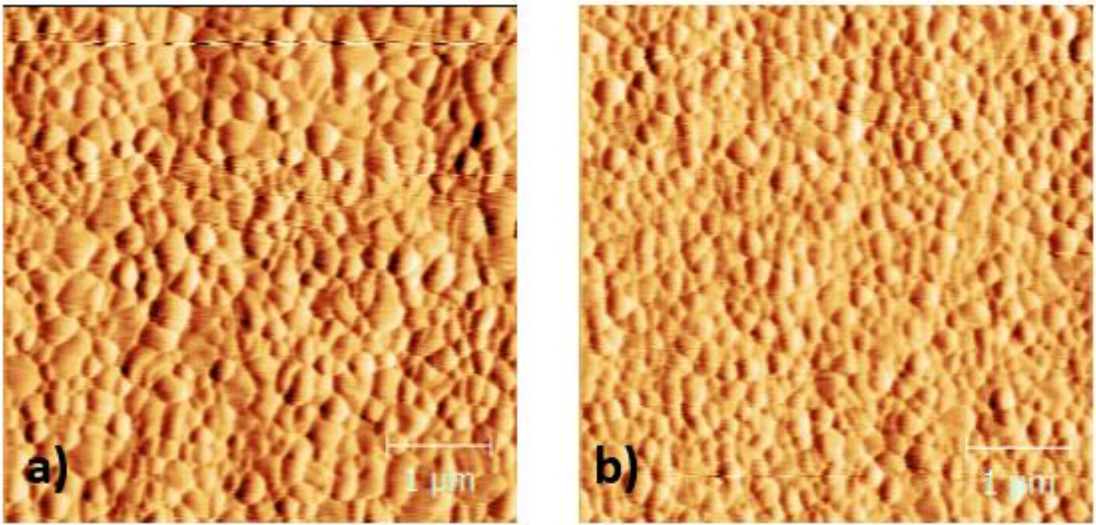

**Figure S1:** AFM measurements in phase contrast of a) ITO/Cu:NiOx/Pvsk, b) ITO/Cu:NiOx/βalanine/Pvsk

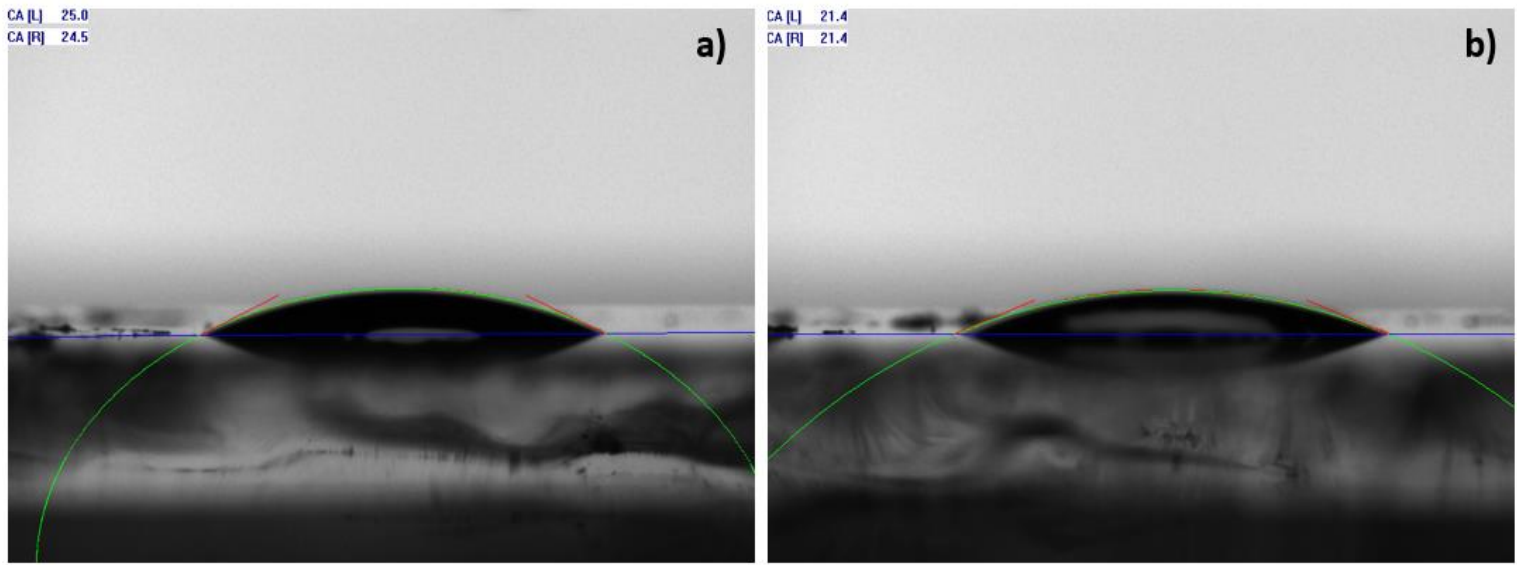

**Figure S2:** Contact angle film measurements of a) Cu:NiOx, b) Cu:NiOx/alanine using γ-butyrolactone:DMSO (7:3) for the perovskite solvent.

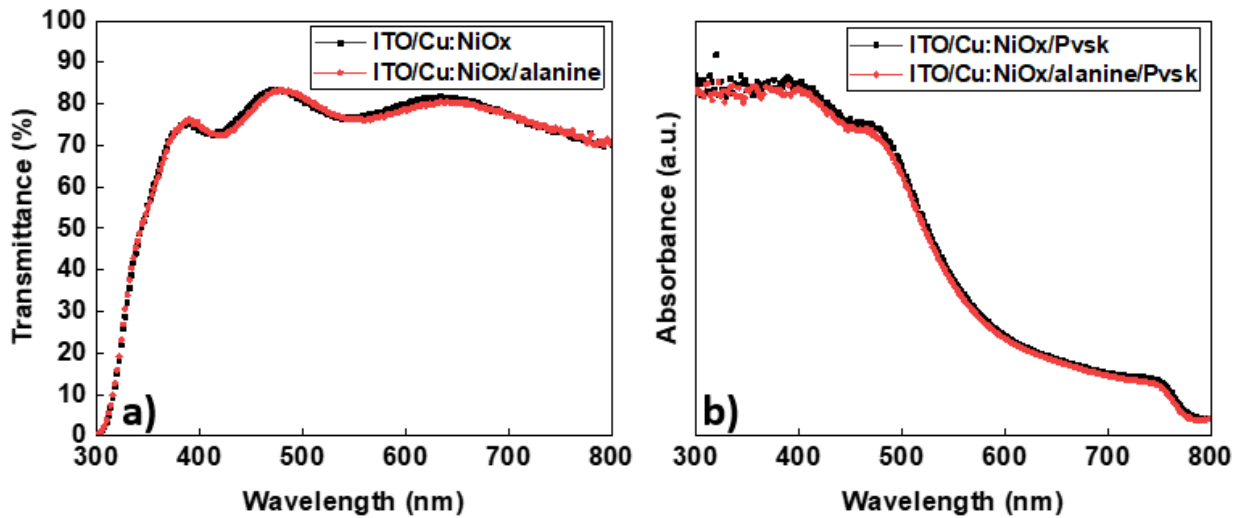

**Figure S3:** a) Transmittance spectra of ITO/Cu:NiOx , ITO/Cu:NiOx/alanine and b) Absorbance spectra ITO/Cu:NiOx/Pvsk, ITO/Cu:NiOx/β-alanine/Pvsk.

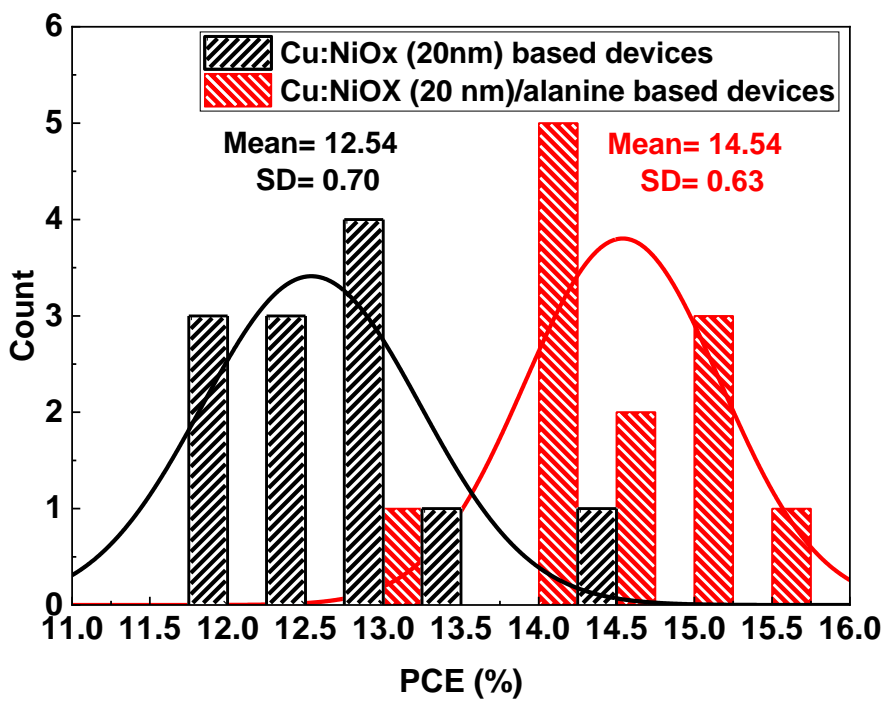

**Figure S4:** Average PCE and SD for the optimized devices based on Cu:NiOx (20 nm)

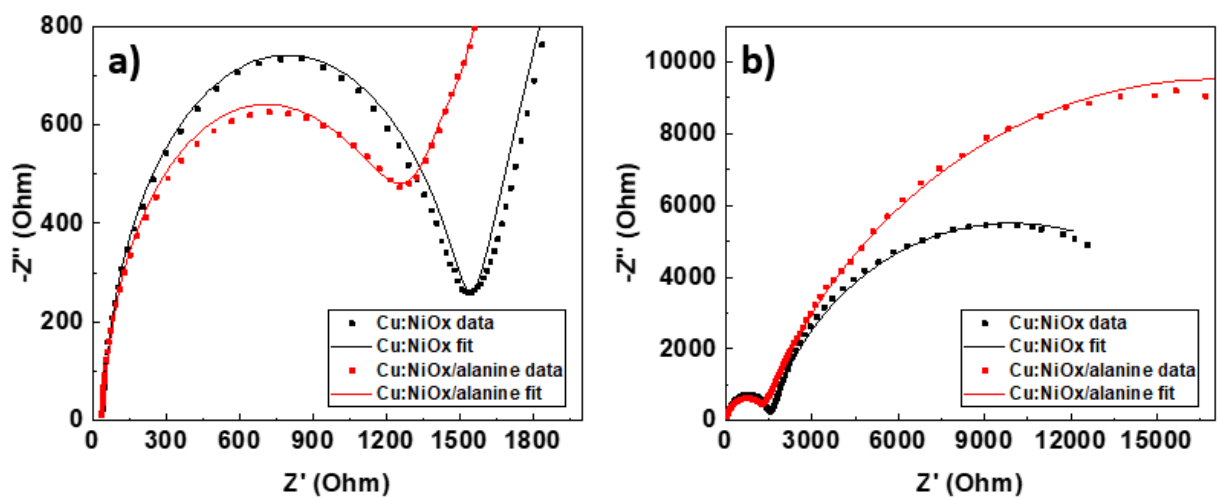

**Figure S5:** a) Nyquist data and fitting for a) $R_{tr}$ and b) $R_{rec}$

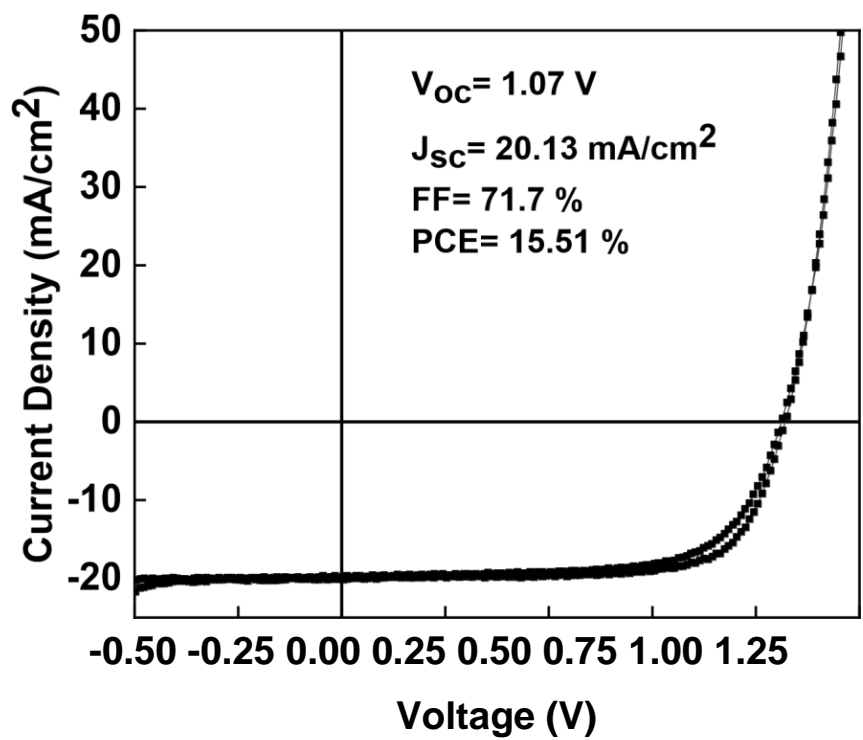

**Figure S6:** J/V characteristic of optimized ITO/Cu:NiOx/β-alanine/Pvsk/PC$_{70}$BM/BCP/Cu